\documentclass[final]{ias2}

\usepackage{graphicx} 
\usepackage{multirow}
\usepackage{array} 

\usepackage{hyperref} 

\begin{document}

\markboth{Photons and dileptons in heavy ion collisions}{T. Sakaguchi}

\title{Photon and dilepton production in high energy heavy ion collisions}

\author[bnlphys]{Takao Sakaguchi} 
\email{takao@bnl.gov}
\address[bnlphys]{Physics Department, Brookhaven National Laboratory, Upton, NY 11973-5000, USA}

\begin{abstract}
The recent results on direct photons and dileptons in high
energy heavy ion collisions, obtained particularly at RHIC and LHC
are reviewed. The results are new not only in terms of the probes,
but also in terms of the precision. We will discuss the
physics learned from the results.
\end{abstract}

\keywords{Photons, Dileptons, RHIC, LHC, QGP}

\pacs{25.75.-q, 25.75.Bh, 25.75.Cj, 25.75.Nq}
 
\maketitle


\section{Introduction}
\begin{figure}[ht]
\centering
\begin{minipage}{16pc}
\includegraphics[width=16pc]{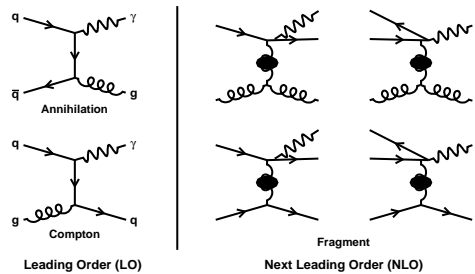}
\end{minipage}
\caption{Production processes of direct photons.}
\label{figBasic}
\end{figure}

\begin{figure}[ht]
\centering
\begin{minipage}{18pc}
\includegraphics[width=18pc]{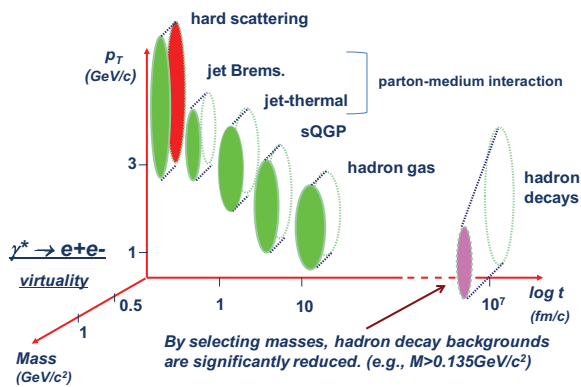}
\end{minipage}
\caption{Manifestation of photons from various sources as a function of formation time and $p_T$.}
\label{figManifest}
\end{figure}

Electromagnetic radiations are an excellent probe for extracting
thermodynamical information of a matter produced in nucleus-nucleus
collisions, as they are emitted from all the stages of collisions with
wide $Q^2$, and don't interact strongly with medium once produced~\cite{ref1}.
They appear in two figures, namely, photons ($\gamma$) that have zero-mass
and virtual photons ($\gamma^*$) that have finite mass. Experimentalists
often refer virtual photons as dileptons since they have been measuring
virtual photons through lepton-pair channels
($\gamma^* \rightarrow ee, \mu\mu$), by which a wide kinematic range in
invariant mass and transverse momentum ($p_T$) can be explored.
Photons are produced through a Compton scattering of quarks and gluons
($qg\rightarrow q \gamma$) or an annihilation of
quarks and anti-quarks ($q\overline{q} \rightarrow g \gamma$) as
leading-order (LO) processes, and the next-to-leading order (NLO) process
is dominated by Bremsstrahlung and fragmentation ($qg \rightarrow qg\gamma$),
as depicted in Figure~\ref{figBasic}.
Their yields are proportional to $\alpha\alpha_{s}$, which are $\sim$40
times lower than those of hadrons that are produced in strong interactions.
Virtual photons are produced by annihilation of quarks and anti-quarks.
The initial hard scattering process that takes place in the beginning of the
collisions produces relatively high $p_T$ photons, often referred as hard
photons. They have same LO and NLO processes, and are called prompt
photons and fragment photons, respectively. The production rate of these
photons are well described by a NLO pQCD calculation~\cite{Gordon:1993qc}.
Photons will be emitted from the hot and dense medium
(quark-gluon plasma: QGP) in high energy nucleus collisions and manifest at
moderate $p_T$(1$<p_T<$3\,GeV/$c$) if the QGP is formed~\cite{Turbide:2003si}.
We often call these photons as thermal photons. The thermal photons
are of interest in exploring thermodynamical nature of the QGP, such
as temperature. One can also obtain the degree of freedom of the system by
combining the temperature with measurement of the energy density of the
system as $g \propto \epsilon/T^4$.
For $p_T<$1\,GeV, the photons are predominantly contributed from hadron gas
state via the processes of
$\pi\pi(\rho) \rightarrow \gamma \rho(\pi)$, 
$\pi K^* \rightarrow  K \gamma$ and etc., which are no longer the
quark-gluon level interaction. We often refer these photons as hadron-gas
photons.
Photons from Compton scattering of hard-scattered partons and partons
in the medium (jet-photon conversion), or Bremsstrahlung of the hard
scattered partons in the medium have been predicted to contribute in the
$p_T$ range of $p_T>$2.5\,GeV/$c$ if QGP is created~\cite{Fries:2005zh}.
Figure~\ref{figManifest} shows a landscape of photon sources as a
function of formation time and $p_T$.
As described above, photons have rich information on the state they
are emitted. However these photons are overwhelmed by the large
background coming from the decay of hadrons.
$\pi^0 \rightarrow \gamma\gamma$ and $\eta \rightarrow \gamma \gamma$
are the major contributors to the background photons ($\sim$95\,\%).
The signal to background ratio at 1-3\,GeV/$c$ is of the order $\sim$10\,\%.
One can understand
the difficulty of the measurement by comparing the uncertainty of the best
$\pi^0$ measurement at RHIC~\cite{Adare:2012wg} which directly relates
to the precision of background determination.
There is a certain probability that photons produced by the same process
acquire virtual mass and decay into lepton-pairs (shown as one another
degree of freedom in Figure~\ref{figBasic}(b)). This process is called
internal conversion process and is different from the virtual photon
production described above. We will explain these photons in detail
in a later section.
There have been several attempts on measuring thermal photons in high
energy nuclear collisions as an evidence of QGP formation.
The first sizable signal was reported by the WA98 experiments at SPS
in 1$<p_T<$3\,GeV/$c$ where a calculation predicts that QGP photons
manifest~\cite{Aggarwal:2000th}.
At that time, the hard photons were not measured because the statistics
ran out at the $p_T$ where the hard photons start arising. Estimating
the hard photon contribution had to rely on a theoretical guidance
that had large ambiguity, therefore the measurement was not able to
exhibit the thermal photon contribution~\cite{Turbide:2003si}.

\begin{figure}[ht]
\begin{minipage}{14.5pc}
\includegraphics[width=14.5pc]{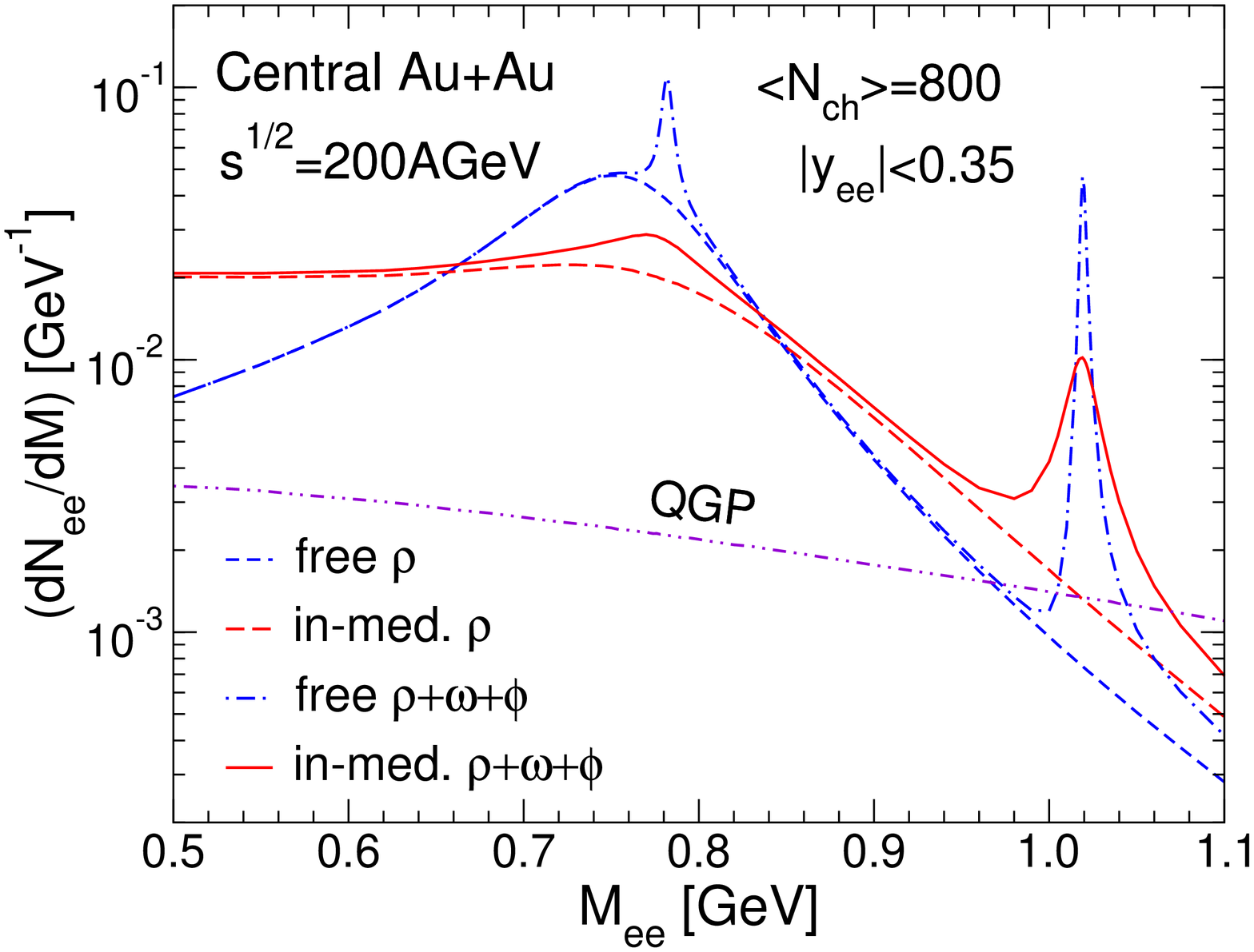}
\end{minipage} \hspace{1pc}%
\begin{minipage}{14.5pc}
\includegraphics[width=14.5pc]{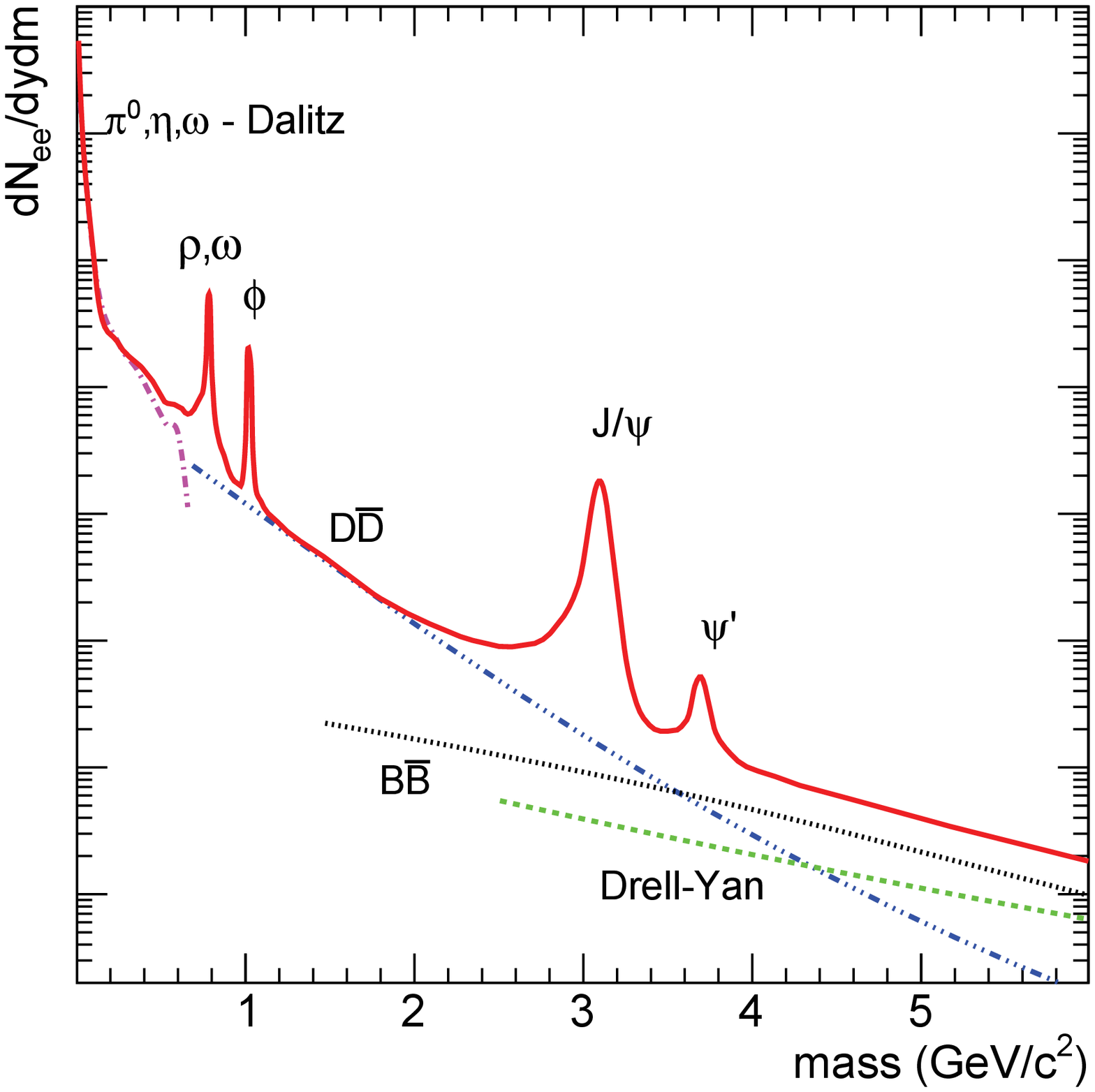}
\end{minipage}
\caption{(a, left) A calculation of dilepton source in low mass region. (b, right) Outline of background components and resonances in dilepton invariant mass spectra.}
\label{figSourceDilepton}
\end{figure}

The source of dileptons depend on their mass and $p_T$ of the measurement.
The low mass region ($m_{ee}<$1\,GeV/$c^2$) at low $p_T$ is predominantly
from the in-medium decay of $\rho$ mesons and/or thermal radiation
as shown in Figure~\ref{figSourceDilepton}(a)~\cite{Rapp:2000pe}.
The same mass region at higher $p_T$ or higher mass region at lower
$p_T$ are from mainly from the thermal radiation.
The dileptons of interest are obtained after subtracting large
combinatoric background arising from Dalitz decays of $\pi^0$ or $\eta$
(e.g., $\pi^0, \eta \rightarrow e^+e^-\gamma$). The outline of the
background components and resonances in dilepton mass spectra is
shown in Figure~\ref{figSourceDilepton}(b)~\cite{Drees:2009xy}.

In this paper, we review the recent measurements of photons and dileptons
from RHIC and LHC experiments and discuss what we have learned from the data.

\section{Hard production of photons and dileptons}
\begin{figure}[ht]
\begin{minipage}{14.5pc}
\includegraphics[width=14.5pc]{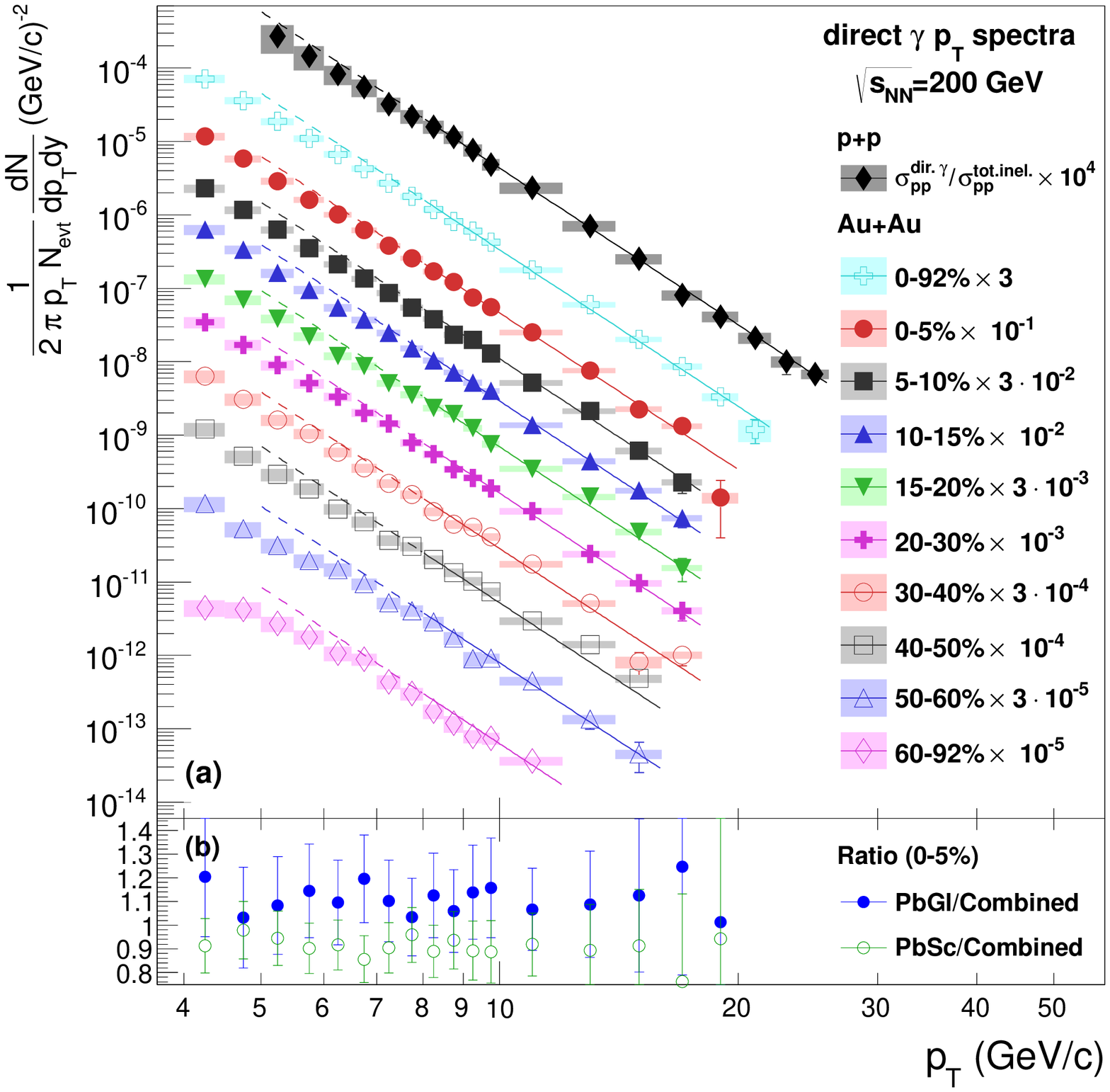}
\end{minipage} \hspace{1pc}%
\begin{minipage}{14.5pc}
\includegraphics[width=14.5pc]{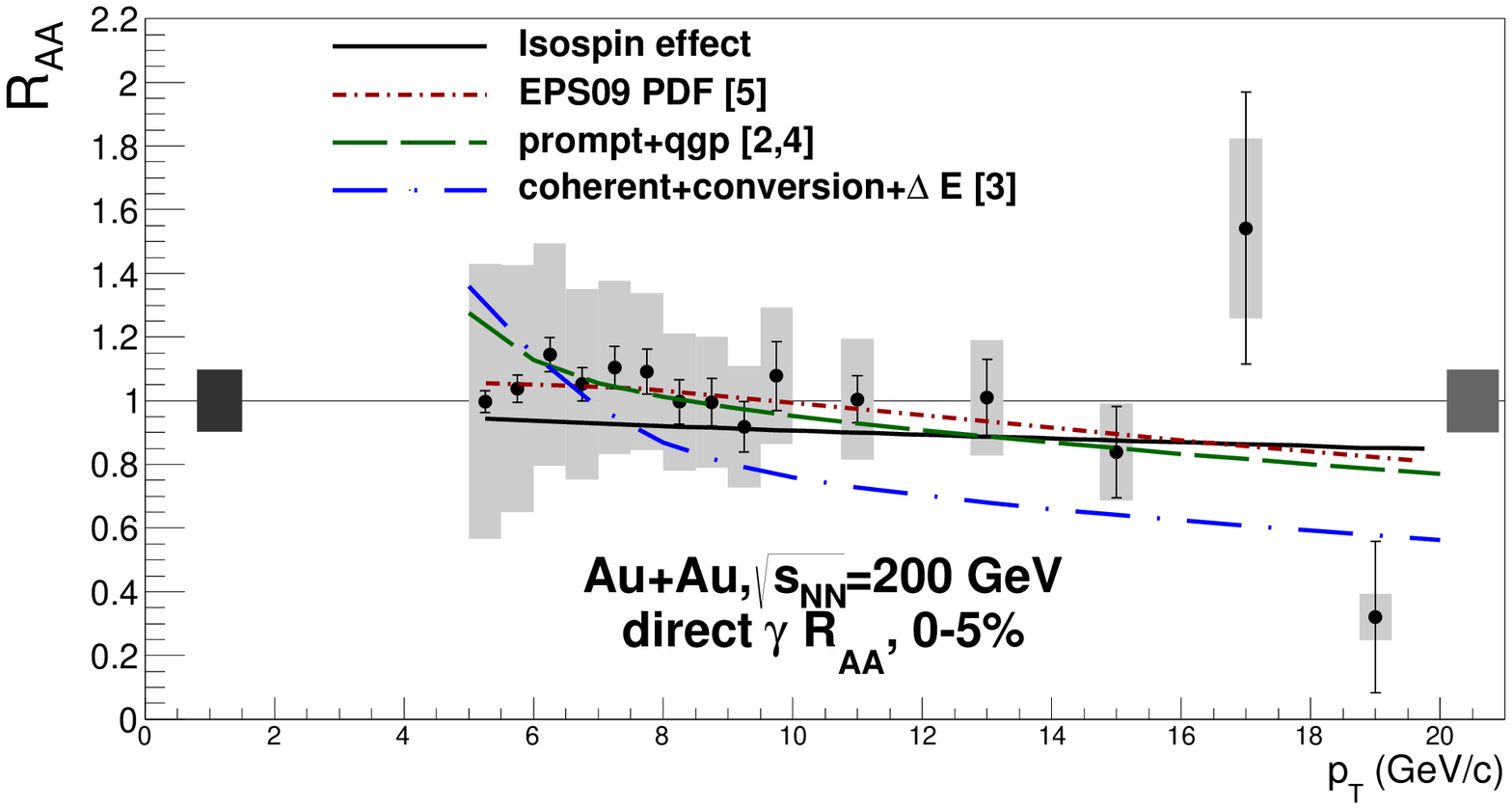}
\end{minipage}
\caption{(a, left) Direct photon $p_T$ spectra in Au+Au collisions at $\sqrt{s_{NN}}$=200\,GeV measured by the PHENIX experiment. (b, right) Nuclear modification factor ($R_{AA}$) for direct photons in 10\,\% central Au+Au collisions~\cite{Afanasiev:2012dg}.}
\label{figPHENIXphotonPPG139}
\end{figure}

\begin{figure}[ht]
\begin{minipage}{14.5pc}
\includegraphics[width=14pc]{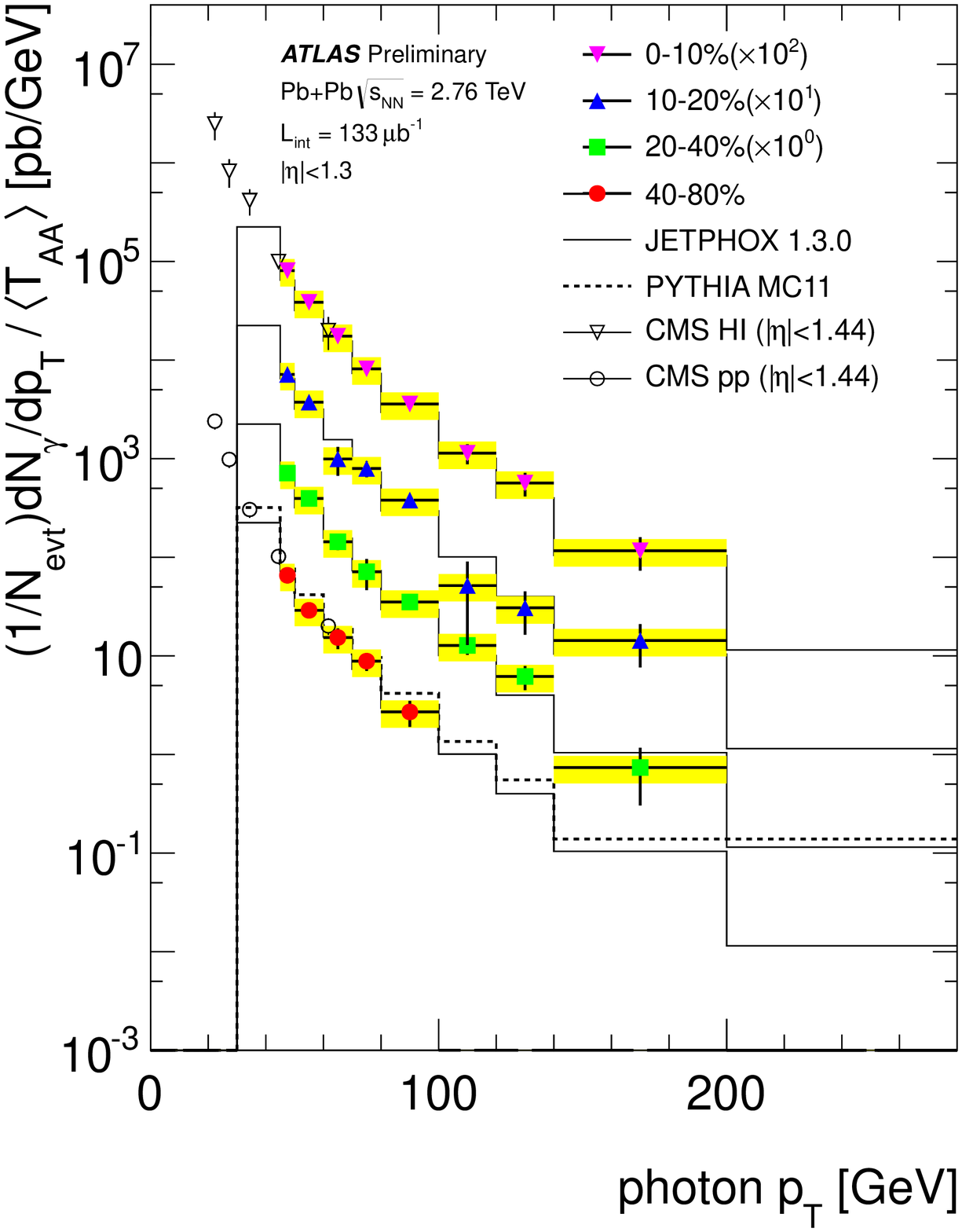}
\end{minipage} \hspace{1pc}%
\begin{minipage}{14.5pc}
\includegraphics[width=14pc]{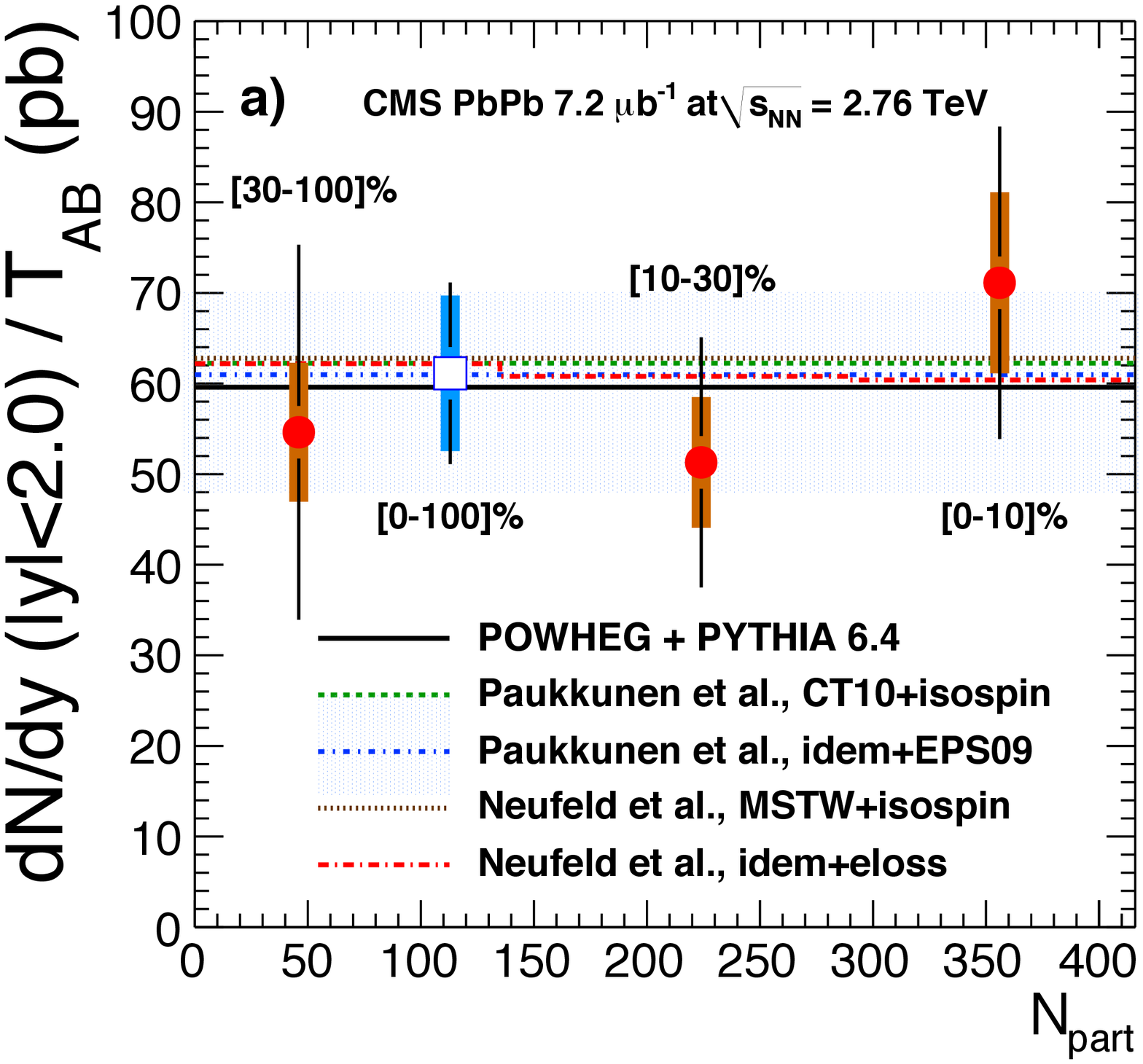}
\end{minipage}
\caption{(a, left) Prompt photon $p_T$ spectra measured in Pb+Pb collisions at $\sqrt{s_{NN}}$=2.76\,TeV by the ATLAS and CMS experiments together with JETPHOX and PYTHIA simulation~\cite{Steinberg:2013laa,Chatrchyan:2012vq}. (b, right) Yield of $Z$-bosons scaled by the nuclear thickness function ($T_{AB}$) as a function of centrality measured by the CMS experiment~\cite{Chatrchyan:2011ua}.}
\label{figLHCphotonAndZ}
\end{figure}

One of the big success in electromagnetic radiation measurements in
relativistic heavy ion collisions is the observation of high $p_T$ direct
photons that are produced in initial hard scattering~\cite{Adler:2005ig}.
Figure~\ref{figPHENIXphotonPPG139}(a) and (b) show the latest direct photon
$p_T$ spectra in Au+Au collisions at $\sqrt{s_{NN}}$=200\,GeV for various
centralities, and the nuclear modification factor ($R_{AA}$) for the
0-10\,\% centrality, respectively~\cite{Afanasiev:2012dg}.
The $R_{AA}$ is consistent with unity within quoted uncertainty, implying
that the photons from Au+Au collisions are consistent with ones expected
from $p+p$ collisions. This result is not very trivial since photons from
hard scattering include both prompt and fragment components. The fragment
photons may be reduced in central Au+Au collisions due to the parton
energy loss. The data is compared with several model predictions. It is
interesting to note that a model that includes reduction of fragment
photon due to energy loss of hard scattered
partons and increase of jet-photon conversion photons (coherent + conversion
+$\Delta E$ ) is not consistent with the data.
The small suppression in $R_{AA}$ seen in the highest $p_T$ is likely due to
the fact that the ratio of the yields in Au+Au to $p+p$ was computed without
taking the isospin dependence of direct photon production into
account~\cite{Arleo:2006xb}.
The LHC heavy ion runs have the cms energy of 2.76\,TeV where the hard
photon production is copious. One can make a photon isolation cut to enrich
the prompt photon component even in heavy ion collisions.
Figure~\ref{figLHCphotonAndZ}(a) shows the isolated photon $p_T$
distributions in Pb+Pb collisions at $\sqrt{s_{NN}}$=2.76\,TeV measured
by the ATLAS experiment at LHC, together with ones by the CMS
experiments~\cite{Steinberg:2013laa,Chatrchyan:2012vq}. The lines show the
expected values from JetPHOX and PYTHIA event generators.
The prompt photon component should be unaffected by the medium created
in heavy ion collisions, and was experimentally confirmed by these
measurements. Considering the fact that neither prompt photons from LHC nor
prompt+fragment photons from RHIC are suppressed, the contribution of
photons from fragmentation and jet-photon conversion could be smaller
than predicted. Several measurements benefited from the fact that these
hard electromagnetic radiations are well under control. For instance,
direct photons are utilized to quantify jet energy loss as they carry
the initial momenta of jets~\cite{Adare:2012qi,Steinberg:2013laa}.
A new hard probe that
became available at the LHC energy is $Z$-bosons. The $Z$-bosons have a peak
mass of $\sim$80\,GeV/$c^2$, and are produced in the medium with a lifetime
of 0.1\,fm/$c$. Therefore, they carry information on the initial states
of the collisions, and decay before they are affected by the medium.
The $Z$-boson yields scaled by nuclear thickness function ($T_{AB}$)
as a function of $N_{part}$ (centrality) published by the CMS experiments
are shown in Figure~\ref{figLHCphotonAndZ}(b)~\cite{Chatrchyan:2011ua}.
The similar results are recently published by the ATLAS
experiments~\cite{Aad:2012ew}. The $Z$-boson yields are found to follow the
scaling of the initial hard scattering process. $Z$-boson is also ideal
for serving as a reference in many similar measurements, such as $Z$-jet
correlations~\cite{Chatrchyan:2013zz}. We point out that $Z$-bosons can
serve as one of the most reliable tools to normalize the dilepton
spectra between $p+p$ and A+A collisions.

\section{Thermal photon production}
\subsection{Spectra}
\begin{figure}[ht]
\centering
\begin{minipage}{20pc}
\includegraphics[width=20pc]{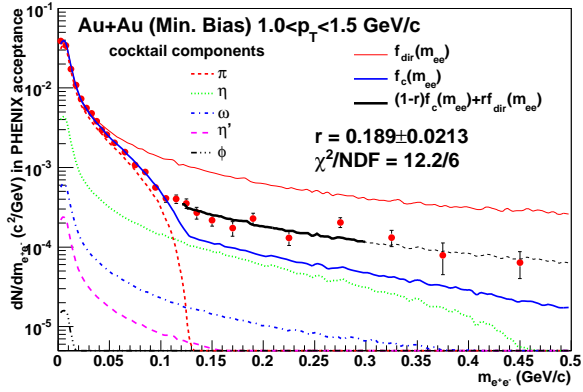}
\end{minipage}
\caption{$e^+e^-$ invariant mass distribution at 1.0$<p_T<$1.5\,GeV/$c$ together with the cocktail calculation of known sources and direct photon internal conversion~\cite{Adare:2008ab}.}
\label{figPHENIXInConmass}
\end{figure}

\begin{figure}[ht]
\begin{minipage}{14.5pc}
\includegraphics[width=14.5pc]{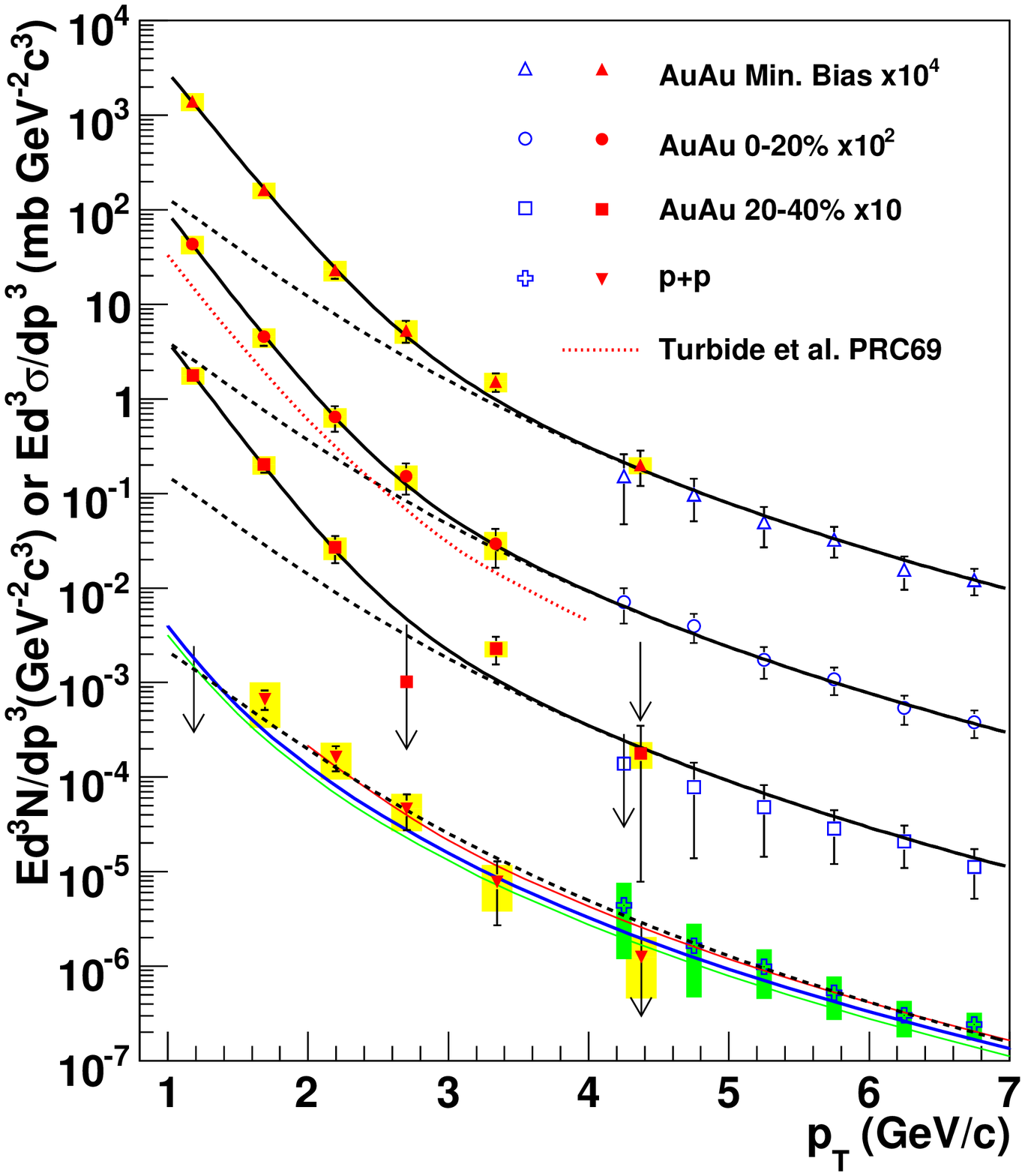}
\end{minipage} \hspace{1pc}
\begin{minipage}{14.5pc}
\includegraphics[width=14.5pc]{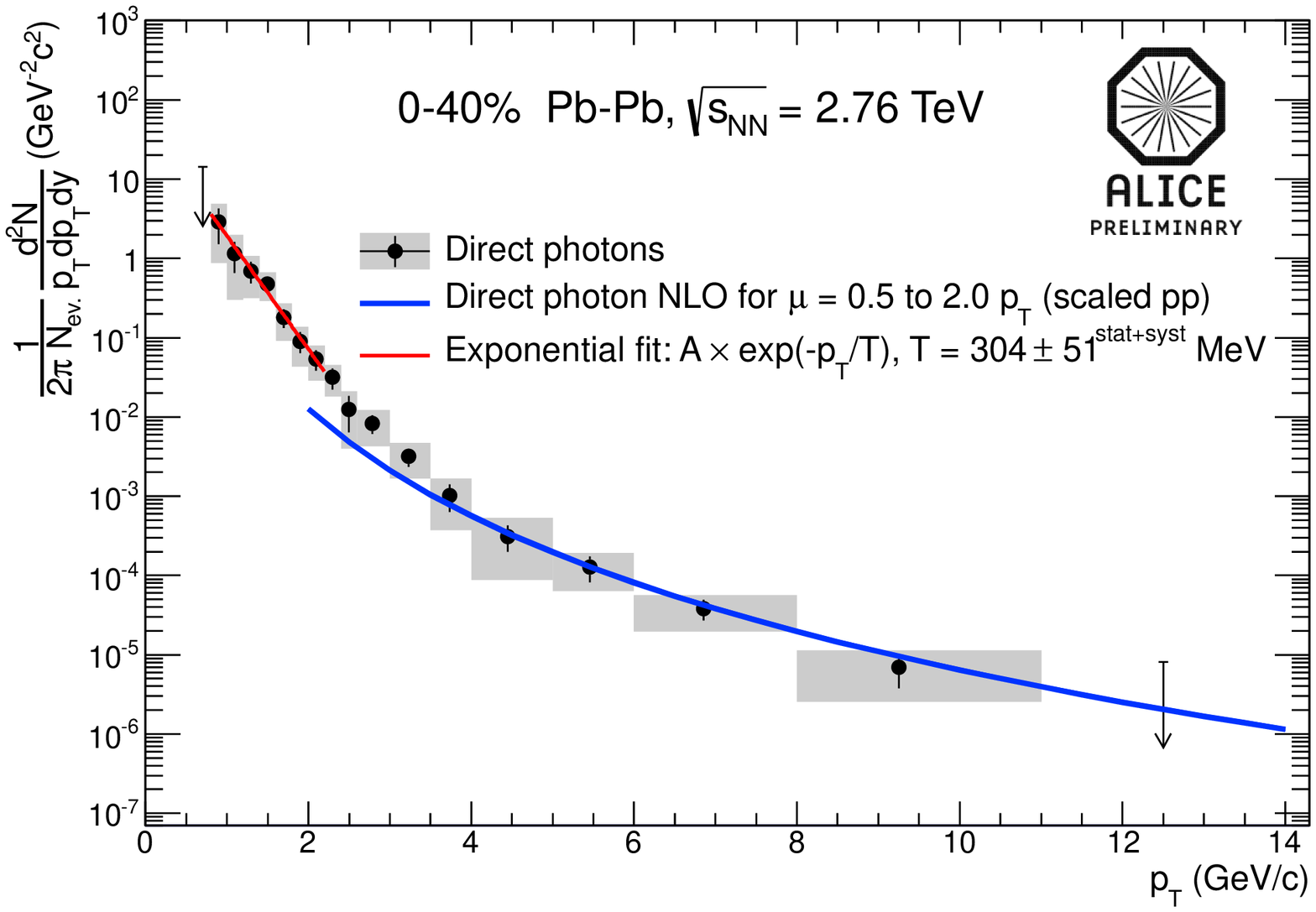}
\end{minipage}
\caption{(a, left) PHENIX results on invariant yields of direct photons in Au+Au collisions at various centralities together with that in $p+p$ collisions. Lines show the $N_{coll}$-scaled $p+p$ fit functions with exponential functions~\cite{Adare:2008ab}. (b, right) ALICE results on invariant yields of direct photons in 2.76\,TeV Pb+Pb collisions~\cite{Wilde:2012wc}}
\label{figPHENIXALICElowpT}
\end{figure}

\begin{figure}[ht]
\centering
\begin{minipage}{20pc}
\includegraphics[width=20pc]{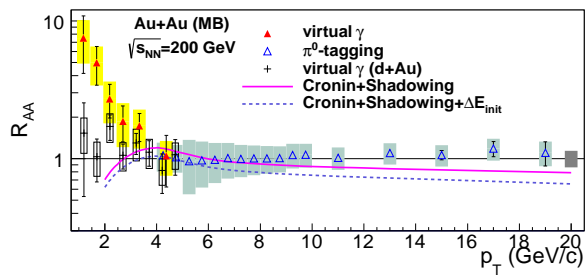}
\end{minipage}
\caption{PHENIX results of direct photon $R_{AA}$ in $d$+Au and Au+Au collisions at $\sqrt{s_{NN}}$=200\,GeV~\cite{Adare:2012vn}.}
\label{figPhotondAu}
\end{figure}

In the conventional real photon measurement, single photons can be observed
after a huge amount of background photons coming from hadron decays
($\pi^0$, $\eta$, $\eta'$ and $\omega$, etc.) are subtracted off from the
inclusive photon distributions. This fact makes it very difficult to look
at the signal at low $p_T$, where thermal photons from QGP manifest.
A breakthrough was made by utilizing internal conversion of
photons~\cite{Adare:2008ab}. Because these photons decay into $e^+e^-$,
one can use measurement technique for dileptons. The relation between
real photon production and the $e^+e^-$ pairs decaying from associated
internal conversion photon production can be described as
follow~\cite{Adare:2009qk}:
\[ \frac{d^2n_{ee}}{dm_{ee}} = \frac{2\alpha}{3\pi}\frac{1}{m_{ee}} \sqrt{1-\frac{4m_e^2}{m_{ee}^2}}\Bigl( 1+\frac{2m_e^2}{m_{ee}^2} \Bigr) S  dn_{\gamma} \]
where $\alpha$ is the fine structure constant, $m_e$ and $m_{ee}$
are the masses of the electron and the $e^+e^-$ pair respectively,
and $S$ is a process dependent factor that goes to 1 when
$m_{ee} \rightarrow 0$ or $m_{ee} \ll p_T$.
This equation also applies to the relation between
the photons from hadron decays (e.g.  $\pi^0\rightarrow \gamma \gamma$)
and the $e^+e^-$ pairs from Dalitz decays ($\pi^0 \rightarrow e^+e^-\gamma$).
For $\pi^0$ and $\eta$, the factor $S$ is given as
$S =|F(m_{ee}^2)|^2 (1-m_{ee}^2/M_h^2)^3$, where $M_h$ is the
meson mass and $F(m_{ee}^2)$ is the form factor. The analysis assumes
that the form factor for direct photons is $F(m_{ee}^2)=1$ similar to
a purely point-like process.
We can select the higher $e+e-$ invariant mass region where $\pi^0$
contribution becomes off. This will eliminate the large background
coming from $\pi^0$ Dalitz decay.
Figure~\ref{figPHENIXInConmass} shows the $e^+e^-$ invariant mass distribution
in minimum bias Au+Au collisions for 1.0$<p_T<$1.5\,GeV/$c$ measured by
the PHENIX experiment~\cite{Adare:2008ab}.
The $e^+e^-$ mass spectra were fit with the function that have terms of
the cocktail calculation of known sources ($e^+e^-$ from various hadron
Dalitz decays) and the direct photon internal conversion:
\[f(m_{ee})=(1-r)f_{c}(m_{ee}) + r~f_{\rm dir}(m_{ee})\]
where $f_{c}(m_{ee})$ is the shape of the cocktail
mass distribution, $f_{\rm dir}(m_{ee})$ is the expected
shape of the direct photon internal conversion.
One can obtain the signal to background ratio for a given mass window
($r$ in the plot) at a given $p_T$.
Using the Kroll-Wada formula~\cite{Kroll:1955zu}, $r$ is associated with
the ratio at zero-mass and thus is converted to the ratio of direct to
inclusive photons:
\[ r = \frac{\gamma^{*}_{\rm dir} 
(m_{ee}>0.15)}{\gamma^{*}_{\rm inc}(m_{ee}>0.15)} \propto \
 \frac{\gamma^{*}_{\rm dir} (m_{ee}\approx 0)}
{\gamma^{*}_{\rm inc}(m_{ee}\approx 0)} \
= \frac{\gamma_{\rm dir}}{\gamma_{\rm inc}} \equiv r_{\gamma} \]
Finally, the direct photon $p_T$ spectra is calculated as
$\gamma_{\rm inc} \times r_{\gamma}$.
Figure~\ref{figPHENIXALICElowpT}(a) shows the direct photon $p_T$ spectra
in Au+Au and $p+p$ collisions as obtained by this procedure.
The spectra were fit with $N_{coll}$-scaled
$p+p$ fit function with exponential function, and the slope parameters of
$\sim$220\,MeV, which is almost independent of centrality, were obtained.
The ALICE experiment recently measured real photons using external
conversion technique, namely, by looking at the conversion of real photons
into $e^+e^-$ in the material of the inner detectors.
Figure~\ref{figPHENIXALICElowpT}(b) shows the direct photon $p_T$ spectrum
measured by the ALICE experiment in 0-40\,\% Pb+Pb collisions at
$\sqrt{s_{NN}}$=2.76\,TeV, together with the exponential fit to the low
$p_T$ region~\cite{Wilde:2012wc}. The ratio of the slope parameter from
PHENIX and ALICE measurements is 1.38. From the published
result on the Bjorken energy density at Pb+Pb top centrality from the CMS
experiment~\cite{Chatrchyan:2012mb} and one at Au+Au from the PHENIX
experiment~\cite{Adler:2004zn}, one can find that the ratio of Bjorken
energy density of the LHC Pb+Pb collisions ($\sim$14\,GeV/fm$^3$) to
that of RHIC Au+Au ($\sim$5.7\,GeV/fm$^3$) is 2.6, which is smaller
than the one expected the ratio of slope parameters (1.38$^4$ = 3.65).
This is because the photons measured experimentally are a sum of the
photons from all the stages from the initial to the final state of
collisions and their slope parameters reflect "average" temperature,
while the energy density is measured at the thermalization.
In order to obtain the temperature at all the stages as well as
the initial energy density, one has to run a hydrodynamical simulation
that has a realistic time profile of the system.
One might ask if the excess of photons over the initial hard scattering
process is due to cold nuclear matter effect (CNM)
such as $p_T$-broadening. PHENIX has recently measured the direct photon
production in $d$+Au collisions to quantify the CNM effect, using the same
internal conversion technique applied to Au+Au collisions.
Figure~\ref{figPhotondAu} shows the $R_{AA}$
of the direct photons in $d$+Au and Au+Au collisions along with a model
on the CNM and parton energy loss effect~\cite{Adare:2012vn}.
It was found that the CNM effect to the direct photon production
is negligible compared to the large excess seen in Au+Au collisions.

\subsection{Flow}
\begin{figure}[ht]
\begin{minipage}{14.5pc}
\includegraphics[width=14pc]{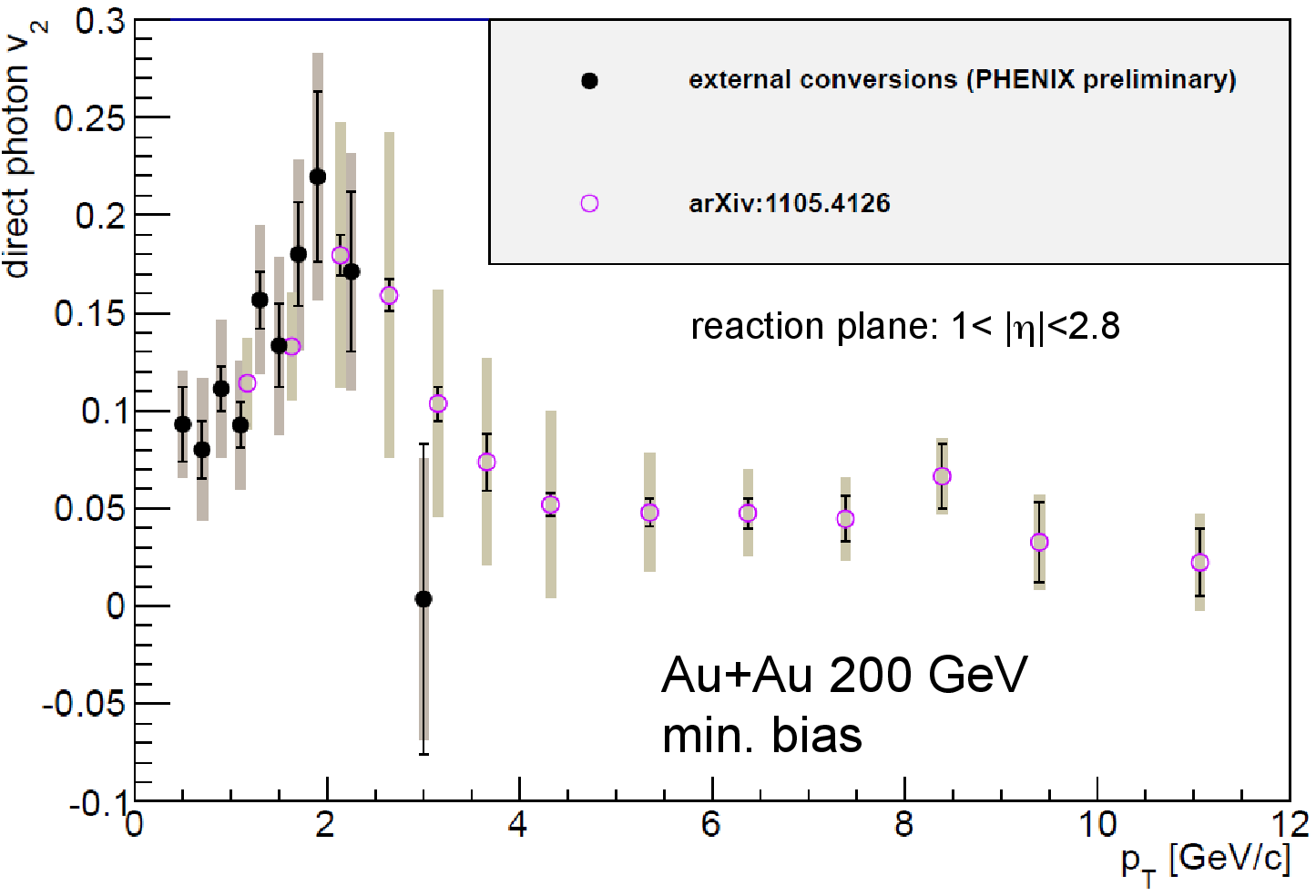}
\end{minipage} \hspace{1pc}%
\begin{minipage}{14.5pc}
\includegraphics[width=14pc]{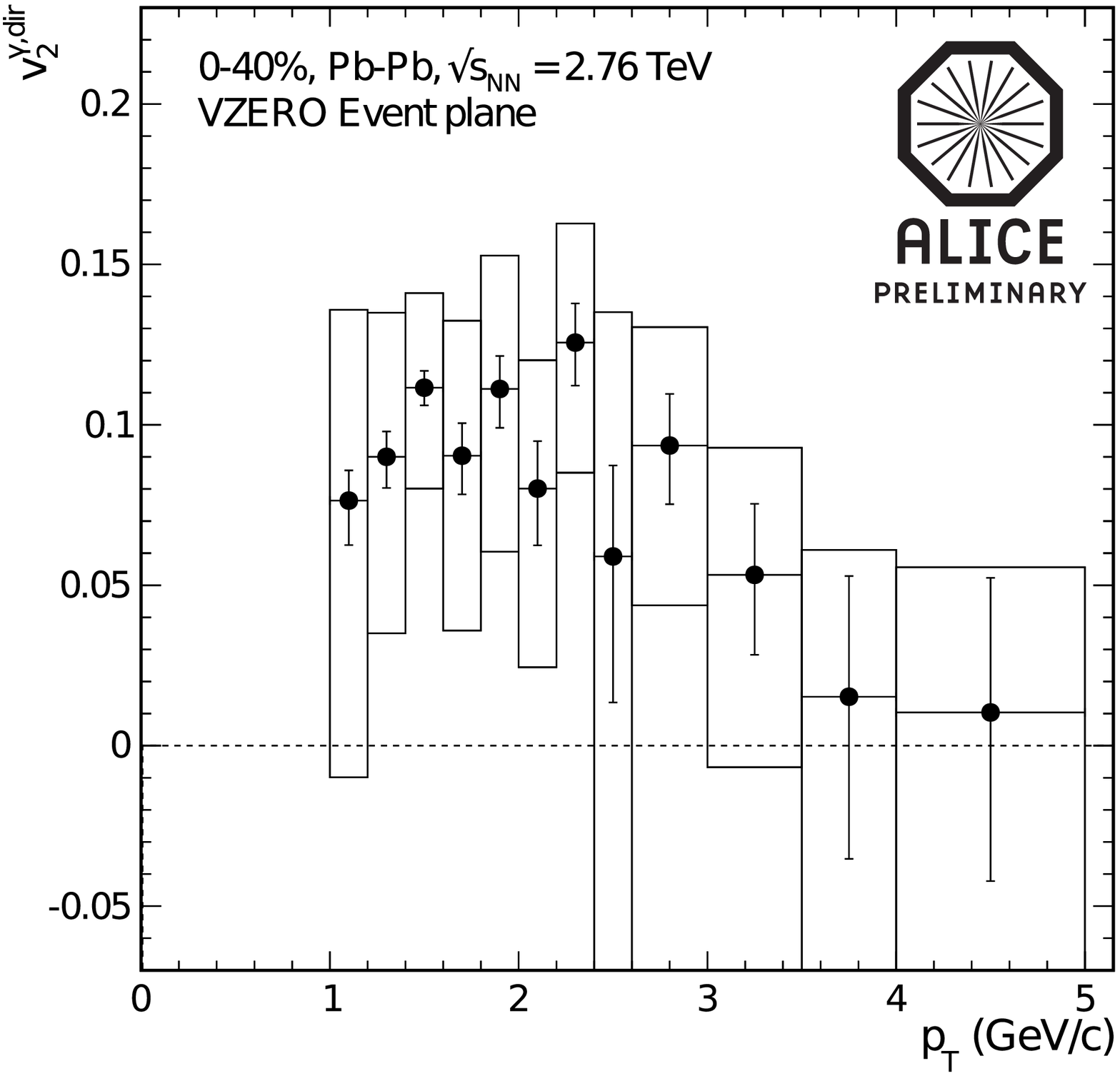}
\end{minipage}
\caption{\label{figPHENIXphotonv2} (a, left) Direct photon elliptic flow ($v_2$) in minimum bias 200\,GeV Au+Au collisions measured by the PHENIX experiment, using internal conversion (marked as arXiv:1105.4126) and external conversion technique~\cite{Adare:2011zr,Tserruya:2012jb}. (b, right) Direct photon $v_2$ in 0-40\,\% 2.76\,TeV Pb+Pb collisions measured by the ALICE experiment using external conversion technique~\cite{Lohner:2012ct}.}
\end{figure}

%
%
After the system reached to local equilibrium, the system proceeds
to hydrodynamic expansion and its interaction with hard scattered
partons, which results in the anisotropic emission of particles.
The magnitude of the anisotropy can be studied from the azimuthal angle
distribution of particles relative to the second order event plane angle
($v_2$, called as elliptic flow). For particles with low $p_T$
($p_T<3$~GeV/$c$), the $v_2$ are understood in terms of pressure-gradient
anisotropy in an initial "almond-shaped" collision zone produced in
non-central collisions.
Recently, a large $v_2$ of particles and its scaling in terms of kinetic
energy have been found for identified charged hadrons at
RHIC~\cite{Adare:2006ti}. It suggested that the system is locally in
equilibrium as early as 0.4\,fm/$c$, and the flow occurs at the partonic
level~\cite{Schenke:2010nt}.
There are predictions that photons also have a collective motion and their
$v_2$ show different signs and/or magnitudes depending on the production
processes~\cite{Turbide:2005bz,Chatterjee:2005de,Turbide:2007mi}. The
observable is useful to disentangle the contributions from various
photon sources in the $p_T$ region where they intermix. The photons
from hadron-gas interaction and thermal radiation may follow the
collective expansion of a system, and give a positive $v_2$. Those
photons produced by jet-photon conversion or in-medium Bremsstrahlung
will increase as the size of the medium to traverse increases and
result in a negative $v_2$. The fragment photons will give positive
$v_{2}$ since larger energy loss of jets is expected in the orthogonal
direction to the event plane. PHENIX has measured the $v_2$ of direct
photons by subtracting
the $v_2$ of hadron decay photons off from that of the inclusive photons,
following the formula below:
\[ {v_2}^{dir.} = (R \times {v_2}^{incl.} -{v_2}^{bkgd.})/(R-1),\ \ \ R = (\gamma/\pi^0)_{meas}/(\gamma/\pi^0)_{bkgd}\]
Here, $R$ is obtained either from internal conversion or external
conversion method, and $v_2^{incl.}$ is obtained from real photons
or their external conversions.
Figure~\ref{figPHENIXphotonv2}(a) shows the direct photon $v_2$ in minimum
bias Au+Au collisions at $\sqrt{s_{NN}}$=200\,GeV measured by the PHENIX
experiment using both internal conversion and external conversion
technique~\cite{Adare:2011zr, Tserruya:2012jb}.
The $v_2$ of direct photons is sizable and positive, and comparable to the
flow of hadrons for $p_T<$3\,GeV/$c$. The ALICE experiments also obtained
the direct photon $v_2$ in 0-40\% Pb+Pb collisions, using external
conversion technique recently~\cite{Lohner:2012ct}. Although the energy
density and the temperature of the two systems are very different,
the $v_2$ at LHC is surprisingly similar to what PHENIX has found at RHIC.

\begin{figure}[ht]
\begin{minipage}{14.5pc}
\includegraphics[width=14pc]{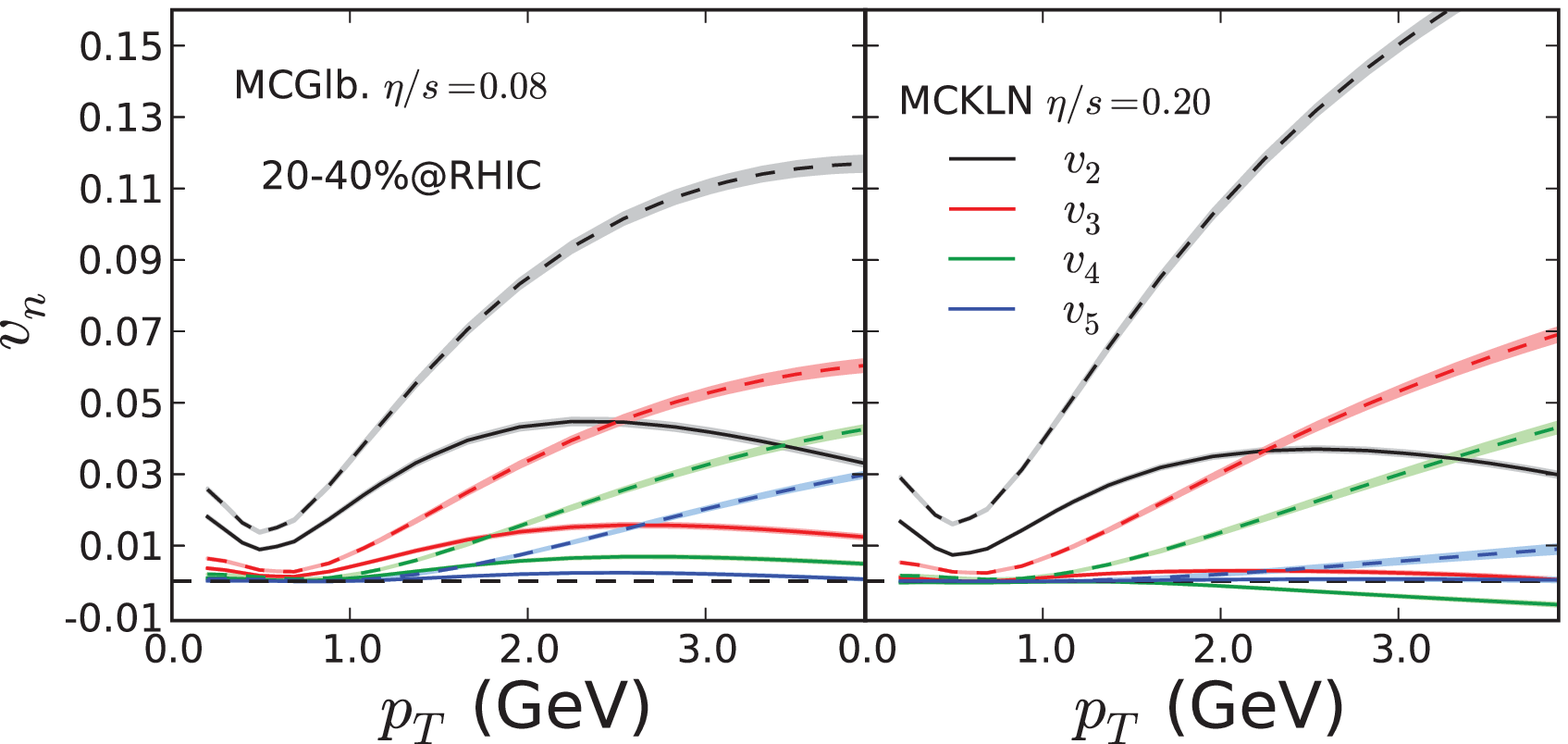}
\end{minipage} \hspace{1pc}%
\begin{minipage}{14.5pc}
\includegraphics[width=14pc]{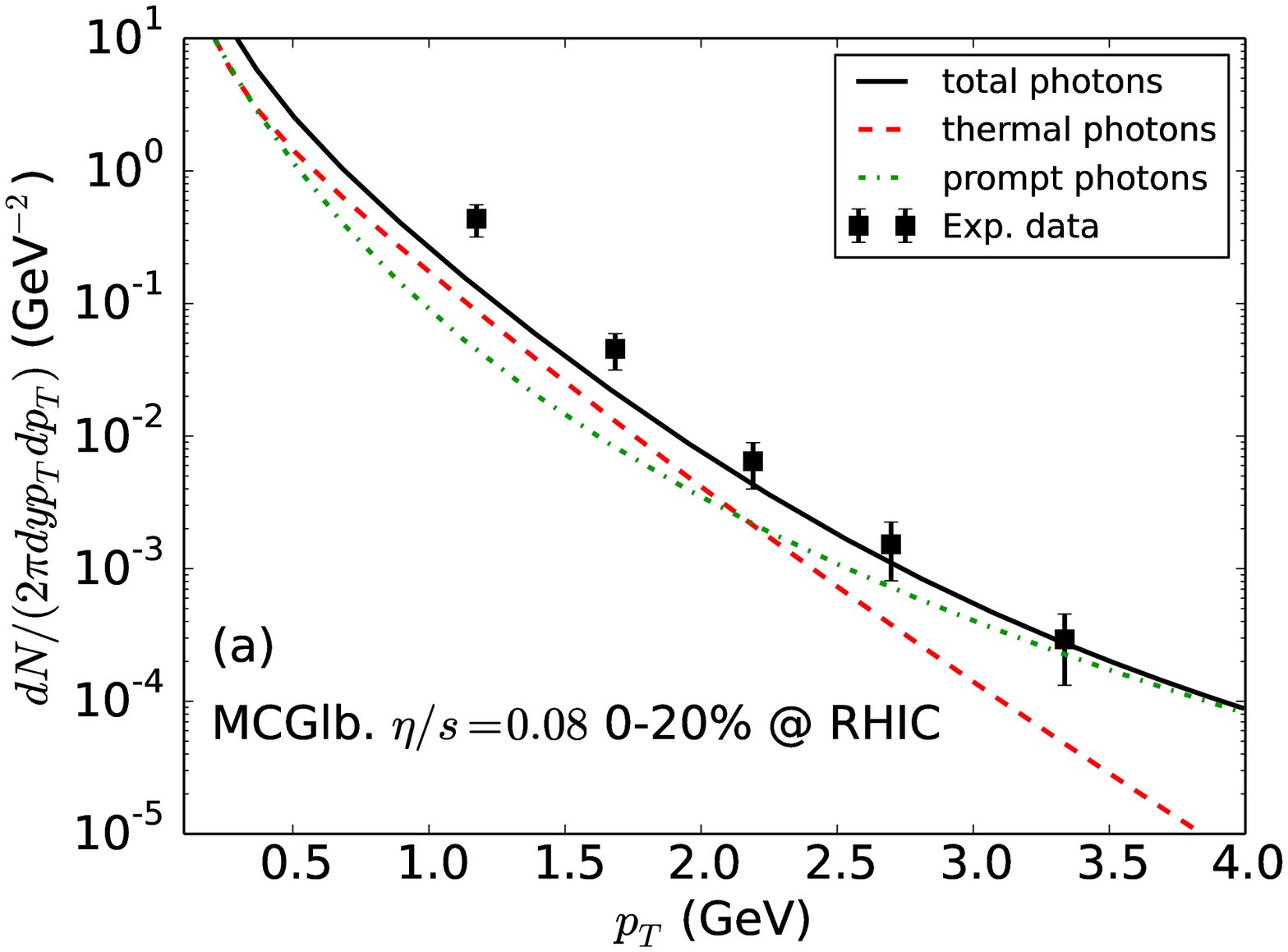}
\end{minipage}
\caption{Direct photon $v_2$ from a theoretical model with including initial state fluctuation and shear viscosity~\cite{Shen:2013cca,Shen:2013vja}. (a, left) $v_2$ from the model calculation, and (b, right) corresponding $p_T$ spectra for MCGlb case.}
\label{figTheoryphotonv2_2}
\end{figure}

There are many models that tried explaining the RHIC result. Several
models predicted the positive flow of the photons assuming the photons
are boosted with hydrodynamic expansion of the system, but the magnitudes
from these models are significantly lower than the
measurement~\cite{Chatterjee:2009qz}.
There are two models that give relatively large flows.
\begin{figure}[ht]
\begin{minipage}{14.5pc}
\includegraphics[width=14pc]{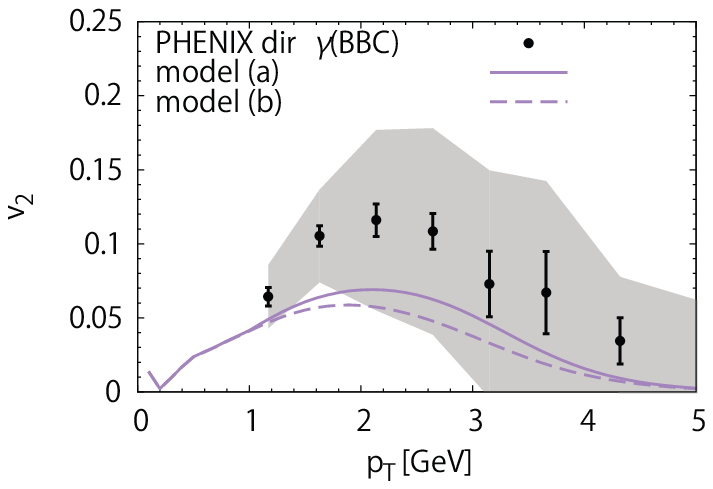}
\end{minipage} \hspace{1pc}%
\begin{minipage}{14.5pc}
\includegraphics[width=14pc]{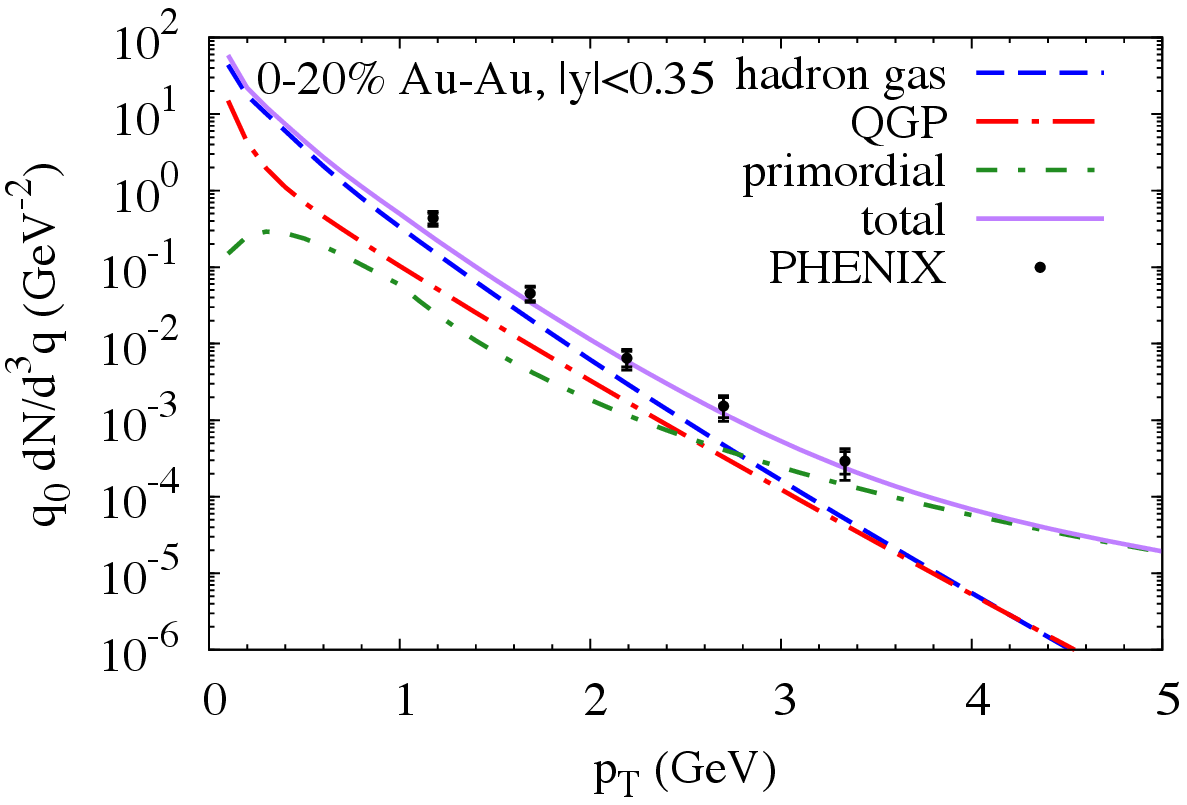}
\end{minipage}
\caption{Direct photon $v_2$ from a theoretical model increasing hadron-gas interaction contribution~\cite{vanHees:2011vb}. (a, left) $v_2$ from the model calculation, and (b, right) corresponding $p_T$ spectra.}
\label{figTheoryphotonv2_1}
\end{figure}
Figure~\ref{figTheoryphotonv2_1} shows a model calculation with increasing
photon contribution from hadron-gas interaction~\cite{vanHees:2011vb}. Since
the hadrons have a large positive flow as we observed, the photons produced
by the interaction with these hadrons result in a large flow. The models (a)
and (b) correspond to two ways of incorporating hard photon contribution,
namely, a pQCD parametrization and a fit to PHENIX p+p data, respectively.
The spectra in low $p_T$ (1$<p_T<$3\,GeV/$c$)
region in this model, where QGP photons are said to dominate, is overwhelmed
by the hadron-gas interaction, and QGP contribution is hardly seen.
Figure~\ref{figTheoryphotonv2_2} shows a model calculation for direct photon
higher order flow ($v_n$) for two initial conditions, Glauber-based (MCGlb)
and CGC-based (MCKLN) conditions, and the corresponding $p_T$ spectrum for
the MCGlb case~\cite{Shen:2013cca,Shen:2013vja}.
The shear viscosity is
increased from $\eta/s$=0.08 to 0.20 when switching from MCGlb to MCKLN,
in order that the model still describes the flow of hadrons. The dashed and
sold lines are before and after viscous corrections are applied on the rates.
The reason that this model gives higher values of $v_n$ for MCKLN is that
the rate of QGP photons are reduced in order to compensate the viscous
entropy production. For a reference, a recent 3+1D hydrodynamic calculation
with a new CGC-inspired initial state (IP-Glasma) gives a good description
of $v_n$ for charged hadrons at RHIC with shear viscosity of
$\eta/s$=0.08~\cite{Gale:2012rq}. This implies that the determination of
$\eta/s$ is significantly affected by initial conditions. Both models
effectively assume a reduction of QGP photons and call for the hadron-gas
photon contribution. Further development at the theory side to explain
the data is clearly deserved. An another model tries to explain the large
flow by the interaction of photons with the strong magnetic field existing
in the non-central collisions~\cite{Bzdak:2012fr}.
The measurement of triangular flow ($v_3$) may be useful to discriminate
the models; for instance, the strong magnetic field scenario gives
$v_3\sim0$ while hydrodynamical expansion scenario gives sizable positive
$v_3$~\cite{Chatterjee:2014nta}.

\section{Low mass dileptons}
\begin{figure}[htb]
\begin{minipage}{14.5pc}
\includegraphics[width=14pc]{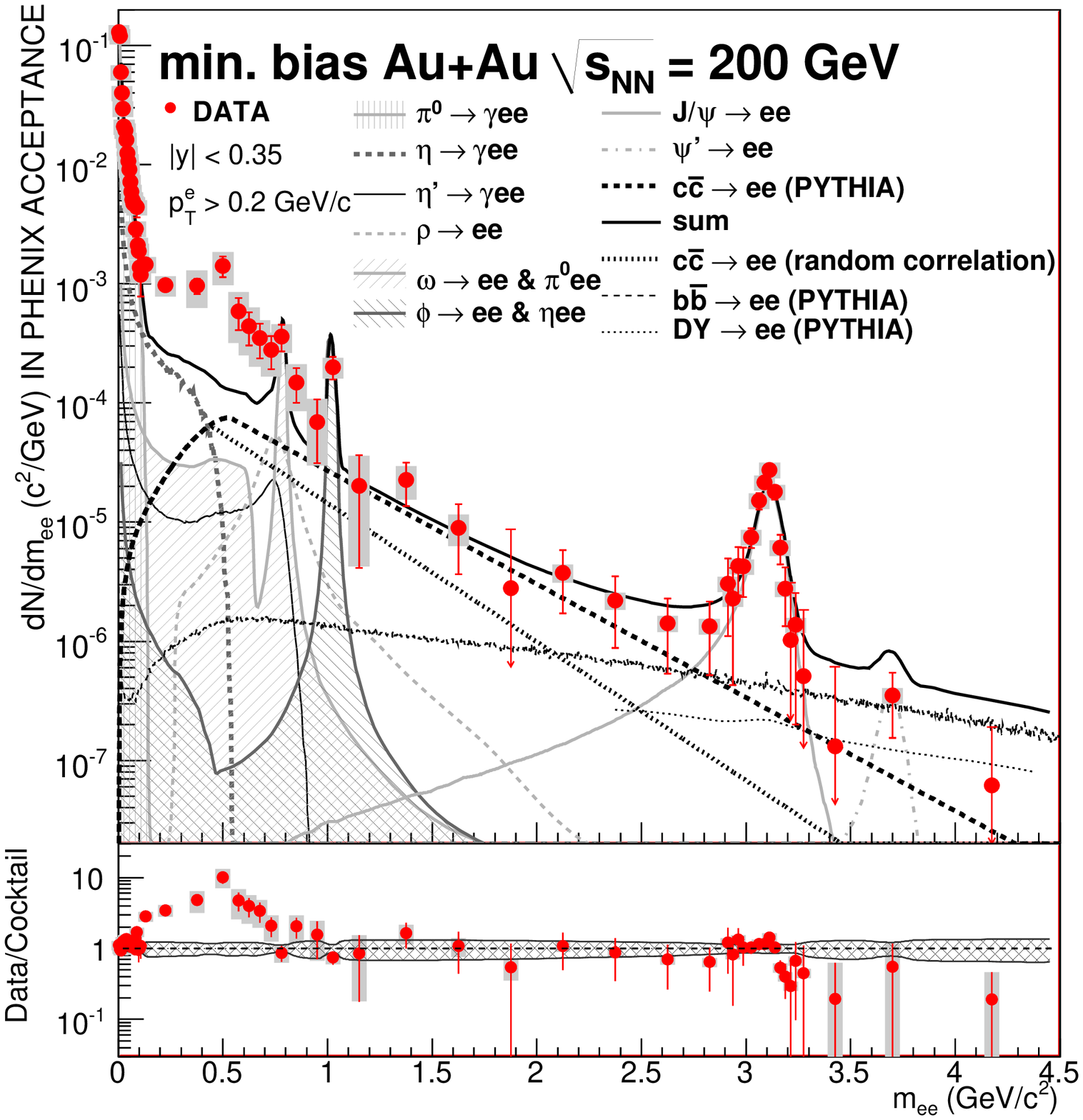}
\end{minipage} \hspace{1pc}%
\begin{minipage}{14.5pc}
\includegraphics[width=14pc]{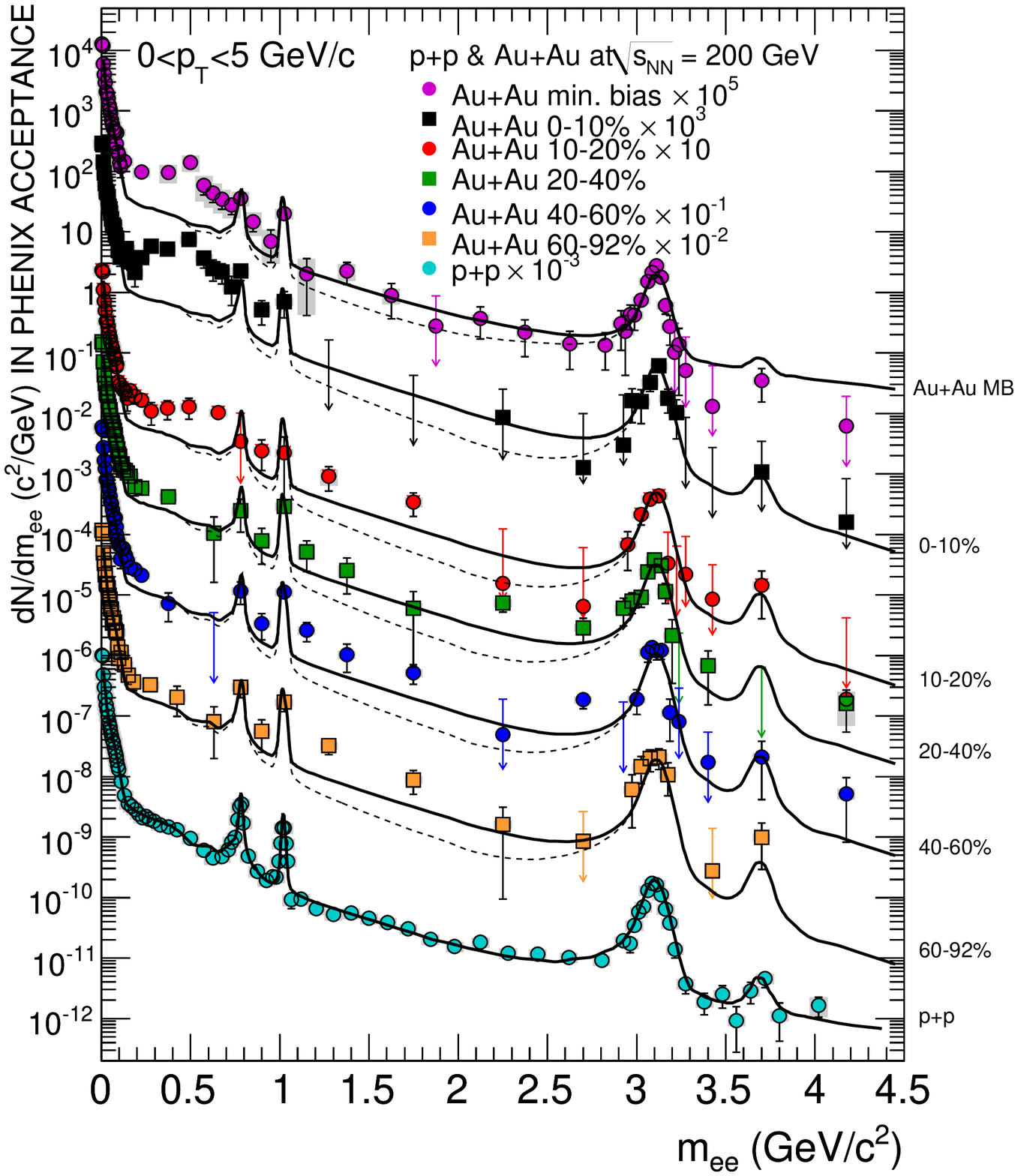}
\end{minipage}
\caption{(a, left) $e^+e^-$ mass spectra in minimum bias 200\,GeV Au+Au collisions measured by the PHENIX experiments together with cocktail calculation of known sources. (b, right) Centrality dependence of $e^+e^-$ mass spectra and corresponding cocktail calculations for the same dataset~\cite{Adare:2009qk}.}
\label{figPHENIXdilepton1}
\end{figure}

The PHENIX experiment has performed the first measurement of the $e^+e^-$
invariant mass spectra in $p+p$ and Au+Au collisions at
$\sqrt{s_{NN}}$=200\,GeV at RHIC as shown in
Figure~\ref{figPHENIXdilepton1}~\cite{Adare:2009qk}.
\begin{figure}[htb]
\centering
\includegraphics[width=23.5pc]{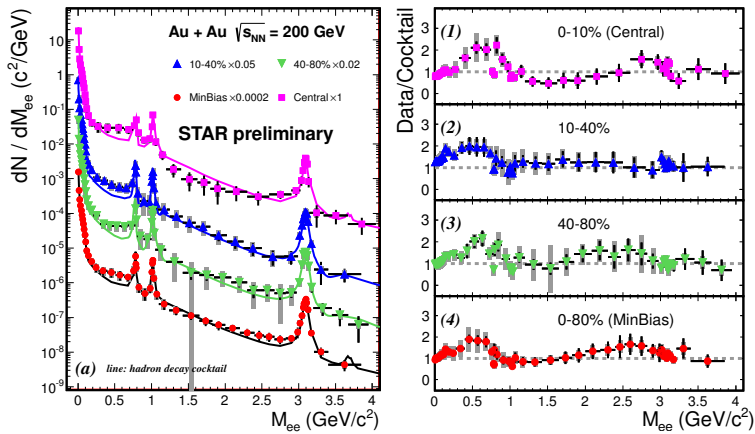}
\caption{(a, left) $e^+e^-$ mass spectra in various centralities in 200\,GeV Au+Au collisions measured by the STAR experiment. (b, right) Ratios of data to cocktail calculation for each centralities~\cite{Geurts:2012rv}.}
\label{figSTARdilepton1}
\end{figure}
\begin{figure}[ht]
\begin{minipage}{14.5pc}
\includegraphics[width=14pc]{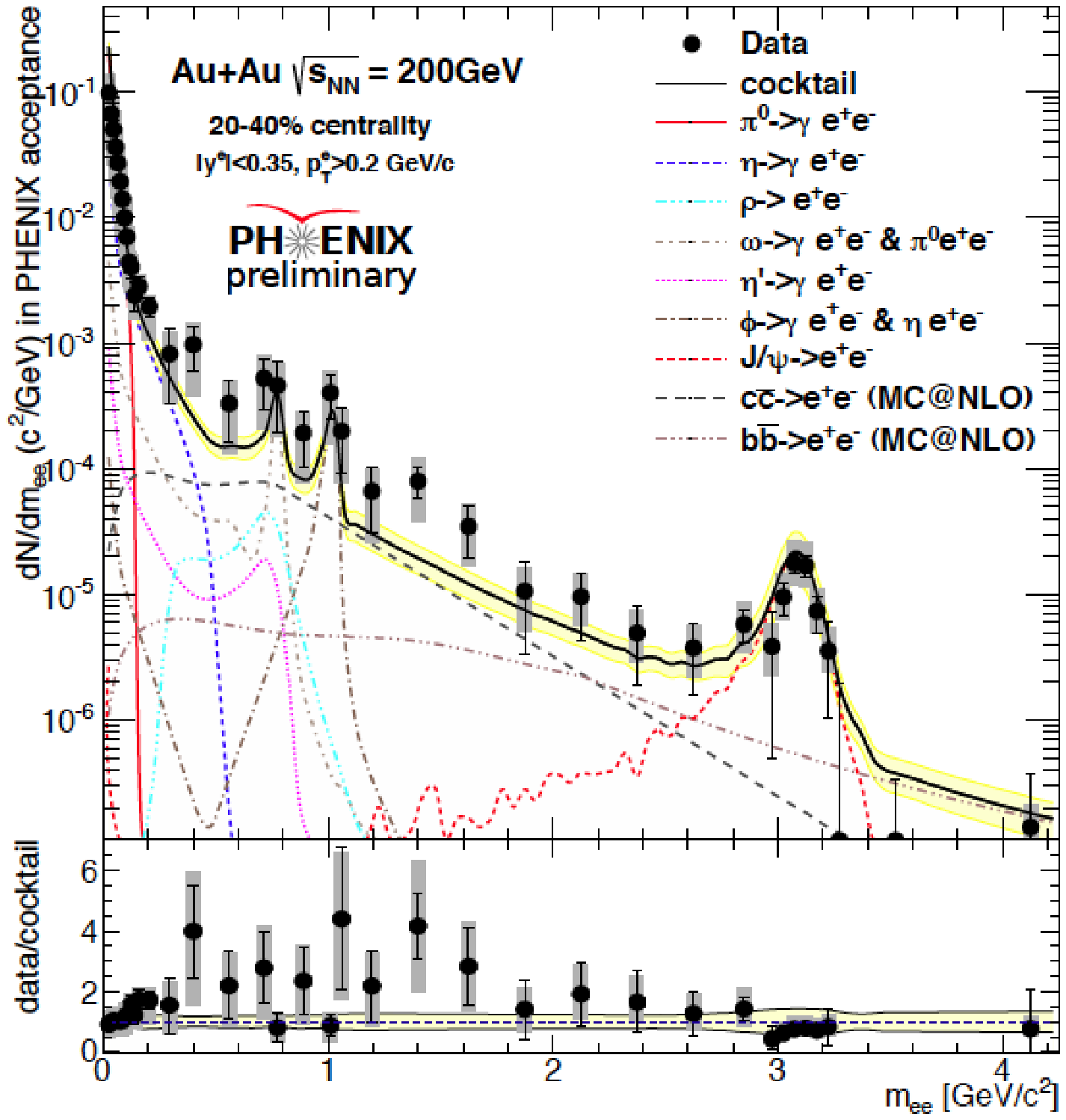}
\end{minipage} \hspace{1pc}%
\begin{minipage}{14.5pc}
\includegraphics[width=14pc]{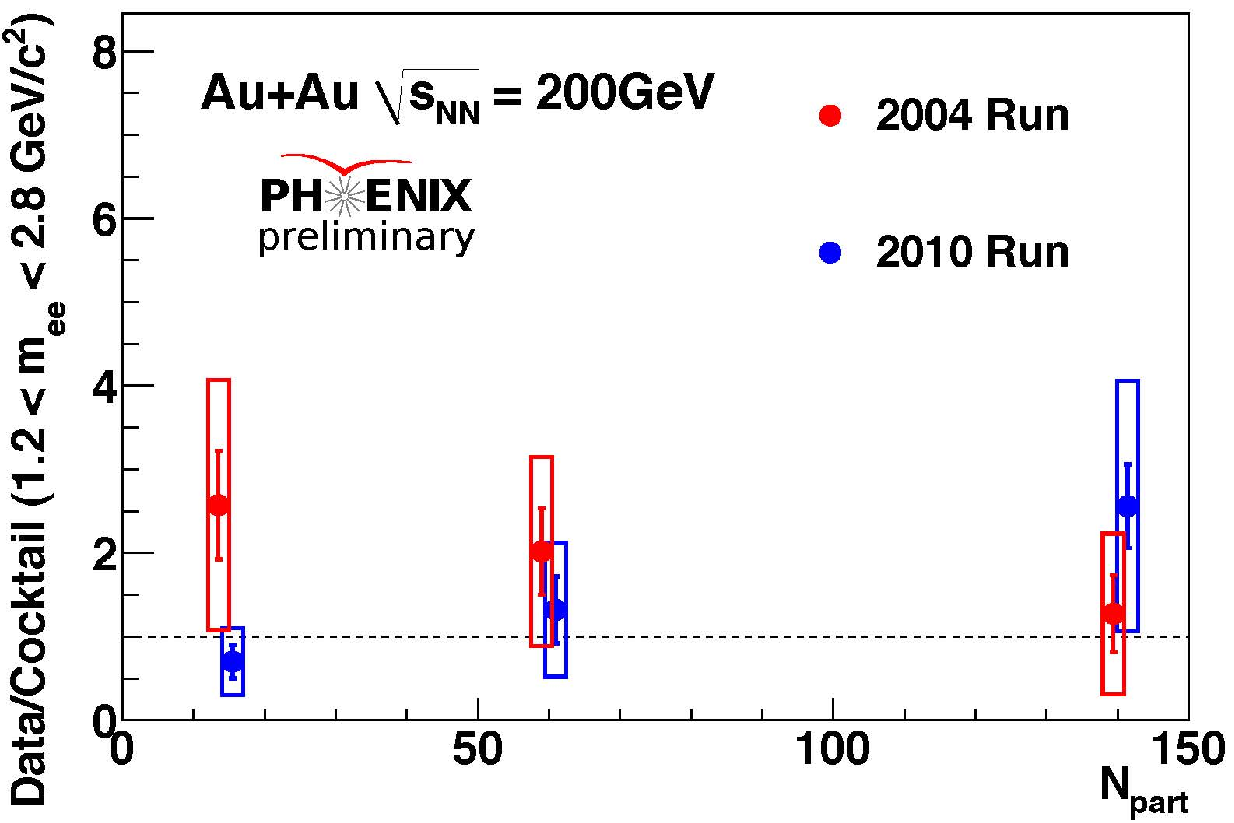}
\end{minipage}
\caption{ (a, left) $e^+e^-$ mass spectra in 20-40\,\% Au+Au collisions measured by the PHENIX experiment in 2010, with a hadron blind detector (HBD) installed. (b, right) Comparison of the ratio of the integrated yields in the LMR to the cocktail calculations in 2004 and 2010 dataset~\cite{Tserruya:2012jb}.}
\label{figPHENIXdilepton2}
\end{figure}
The left panel shows the minimum bias mass spectra with hadronic cocktail
components. The right panel shows the centrality dependence of mass spectra
with cocktail calculations. The $p+p$ results are well reproduced by the
cocktail of known sources of $e^+e^-$, whereas the Au+Au data show a strong
enhancement at low mass region (LMR, $0.15\!<M_{ee}\!<\!0.75$\,GeV/$c^{2}$)
compared to cocktail calculations. The enhancement for the 0-10\,\% central
collisions is a factor of
$7.6 \pm 0.5^{stat} \pm 1.3 ^{syst} \pm 1.5 ^{model}$ and that for minimum
bias collisions is $4.7 \pm 0.4^{stat} \pm 1.5 ^{syst} \pm 0.9 ^{model}$,
respectively.
The STAR experiment also recently obtained $e^+e^-$ mass spectra in $p+p$
and Au+Au collisions at $\sqrt{s_{NN}} = 200$\,GeV as shown in
Figure~\ref{figSTARdilepton1}~\cite{Geurts:2012rv,Adamczyk:2013caa}.
In the same LMR, the enhancement for 0-10\,\%
central collisions is $1.72\pm0.10\pm0.50$, and that for minimum bias
collisions is $1.53\pm0.07\pm0.41$, respectively. Clearly, there is
a discrepancy in the magnitudes of the excess by a factor of 3-4 between
two experiments. PHENIX has installed a hadron blind detector (HBD) in 2010
run in order to reduce systematic uncertainty of the measurement and also
to confirm the previous result. The HBD is a Cherenkov detector with CF4 gas
and rejects the $e^+e^-$ tracks from photon conversions and Dalitz
decay of $\pi^0$'s, which are major background in LMR measurement, by
looking at the opening angle of pair tracks in a magnetic field free
region~\cite{Anderson:2011jw}. Figure \ref{figPHENIXdilepton2}(a) shows the
$e^+e^-$ mass spectra measured in 20-40\,\% Au+Au collisions in the 2010
run with the HBD~\cite{Tserruya:2012jb}.
The analysis for the most central collisions are still in progress.
PHENIX measurements of 2004 and 2010 have several differences in both
data (magnetic field and detector material budget), and cocktail
calculations (MC@NLO is used for charm contribution estimate in 2010).
PHENIX has made a comparison of the data/cocktail ratio for the
integrated yield in LMR obtained in the 2004 and 2010 runs for
the three centrality bins as shown in Figure~\ref{figPHENIXdilepton2}(b).
It is seen that the two runs give consistent results within uncertainties.
It should be noted that the previous PHENIX data showed
most of the excess is seen in most central events (0-20\,\%), whose
analysis is still in progress. The discrepancy between PHENIX and
STAR results persists until PHENIX comes up with the new result with HBD.
As for the STAR data, some concern is put on the fact that the cocktail
calculation in $p+p$ collisions over-predicts
data points in the LMR, though they are still consistent
within systematic errors. If one correct for the over-prediction,
it is possible STAR and PHENIX see the same amount of excess.
Further effort to understand the discrepancy is deserved.

As measured for photons, the measurement of dilepton flow is useful.
There is a radial flow measurement of dileptons by the NA60
experiment~\cite{Arnaldi:2007ru}, and an elliptic flow measurement by
the STAR experiment~\cite{Cui:2013hla,Adamczyk:2014lpa} that could help
disentangling the source of the dileptons that contribute to this particular
mass region. A theory study is also in progress~\cite{Vujanovic:2013jpa}.

\subsection{Energy dependence of LMR dilepton production}
\begin{figure}[ht]
\centering
\includegraphics[width=23.5pc]{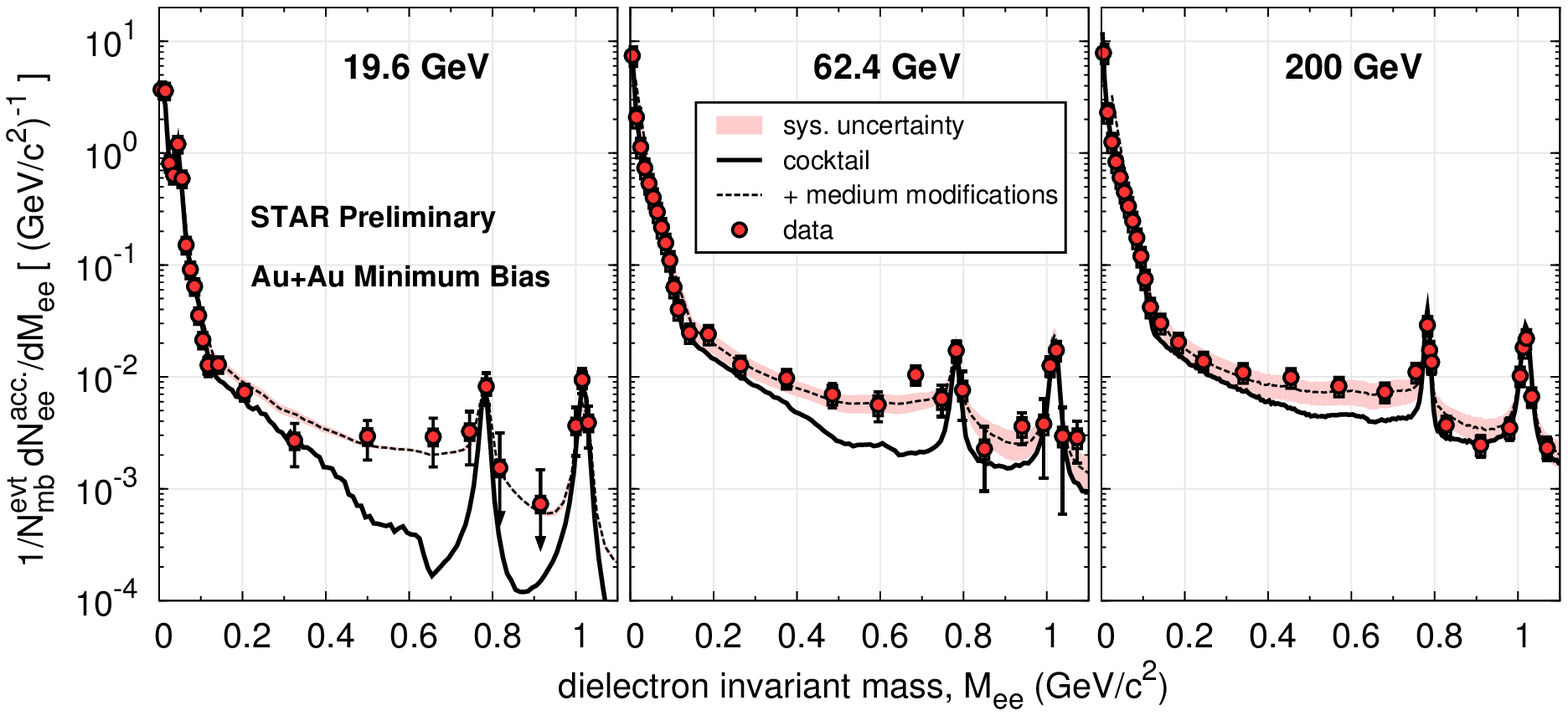}
\caption{$e^+e^-$ mass spectra measured by the STAR experiment in 19.6, 62.4 and 200\,GeV Au+Au collisions, together with cocktail calculation and a theoretical model assuming in-medium modification of $\rho$ spectral function~\cite{Geurts:2012rv}.}
\label{figSTARdilepton2}
\end{figure}
The excess in LMR has been explained by various models including in-medium
broadening of the $\rho$ mesons or their mass shift~\cite{Rapp:1999us}.
The measurement of the energy dependence of the excess may provide a
discrimination of these models.
The STAR experiment recently came up with the $e^+e^-$
invariant mass spectra for minimum bias Au+Au collisions at
$\sqrt{s_{NN}}=$ 19.6, 62.4, and 200\,GeV as shown
in Figure~\ref{figSTARdilepton2}~\cite{Geurts:2012rv}.
STAR observed a qualitatively similar excess as observed by the CERES
measurements
in the Pb$+$Au at $\sqrt{s_\mathrm{NN}}=$17.2\,GeV~\cite{Agakichiev:2005ai}.
In each of the three panels the hadron cocktail simulation includes
contributions from Dalitz decays, photon conversions (19.6\,GeV only),
and the dielectron decay of the $\omega$ and $\phi$ vector mesons. The
cocktail simulations purposely exclude contributions from $\rho$ mesons.
Instead, these are explicitly included in the model calculation by
Rapp~\cite{Rapp:1999us} which involve in-medium modifications of the
$\rho$ meson spectral shape in the isentropic fireball evolution.
The LMR enhancement measured by STAR are consistently agreeing with these
model calculations within the quoted errors. One should be careful on
taking the absolute magnitude of the excess in all energies, provided
that there is still the issue of inconsistency between STAR and
PHENIX. Nonetheless, it is interesting to see that the relative
change as a function cms energy is well described by this model
calculation. If it is the case, the flow of this mass region may exhibit
the $KE_T$ scaling with assuming they are all $\rho$.

\section{Summary}
The recent results on direct photons and dileptons in high
energy heavy ion collisions, obtained particularly at RHIC and LHC
are reviewed. The results are new not only in terms of the probes,
but also in terms of the precision. Much progress has been made
in understanding high $p_T$ direct photons as well as $Z$-bosons with the
latest RHIC and LHC results. The soft single photons have been studied
down to lower $p_T$ using internal conversion and external conversion
technique at RHIC and LHC, and exhibit the average temperature of
the system. A large flow of soft photons was also observed both at RHIC and
LHC, which are not explained by models considering QGP only so far.
The low mass dilepton excess was observed at the PHENIX and STAR
experiments at RHIC in Au+Au collisions, but are not quantitatively
agreeing each other. The flow of dilepton is as important as the one
of photons, and experiments should make an effort to improve the
measurement.
The STAR results on dileptons at several cms energies shows that
the excess in LMR is consistent with in-medium modification
of the $\rho$ meson spectral function. If it is the case, the flow of
this mass region may exhibit the $KE_T$ scaling with assuming they
are all $\rho$. This is an interesting topic to explore.

\bibliographystyle{pramana}

\end{document}